\documentclass[12pt,preprint]{aastex}

\begin{document}

\title{Collapse and Fragmentation of Magnetic Molecular Cloud Cores with 
the Enzo AMR MHD Code. I. Uniform Density Spheres}

\author{Alan P.~Boss and Sandra A.~Keiser}
\affil{Department of Terrestrial Magnetism, Carnegie Institution for
Science, 5241 Broad Branch Road, NW, Washington, DC 20015-1305}
\email{boss@dtm.ciw.edu}

\begin{abstract}

Magnetic fields are important contributers to the dynamics of collapsing 
molecular cloud cores, and can have a major effect on whether collapse 
results in a single protostar or fragmentation into a binary or multiple 
protostar system. New models are presented of the collapse of magnetic 
cloud cores using the adaptive mesh refinement (AMR) code Enzo2.0. 
The code was used to calculate the ideal magnetohydrodynamics 
(MHD) of initially spherical, uniform density and rotation clouds with  
density perturbations, i.e., the Boss \& Bodenheimer (1979) standard 
isothermal test case for three dimensional (3D) hydrodynamics (HD) codes. 
After first verifying that Enzo reproduces the binary fragmentation expected 
for the non-magnetic test case, a large set of models was computed with 
varied initial magnetic field strengths and directions with respect to 
the cloud core axis of rotation (parallel or perpendicular), density
perturbation amplitudes, and equations of state. Three significantly
different outcomes resulted: (1) contraction without sustained collapse,
forming a denser cloud core, (2) collapse to form a single protostar 
with significant spiral arms, and (3) collapse and fragmentation
into binary or multiple protostar systems, with multiple spiral arms.
Comparisons are also made with previous MHD calculations of similar clouds
with barotropic equations of state. These results for the collapse of 
initially uniform density spheres illustrate the central importance of both 
magnetic field direction and field strength for determining the outcome of 
dynamic protostellar collapse. 

\end{abstract}


\section{Introduction}

 Observations of dark cloud cores have shown that magnetic fields are often an 
important contributer to cloud support against collapse for densities in the 
range of $10^3 -10^4$ cm$^{-3}$ (e.g., Troland \& Crutcher 2008; Crutcher 2012). 
As a result, it has become increasingly routine for theoretical models of 
protostellar system formation to include magnetic fields. Boss (1997) used 
a pseudo-MHD 3D gravitational hydrodynamics code with radiative transfer in 
the Eddington approximation to study the collapse and fragmentation of 
magnetic clouds, including the effects of ambipolar diffusion. More recently, 
true MHD codes have become the standard. E.g., Banerjee \& Pudritz (2006) 
considered the collapse of a magnetized cloud core, finding that the cloud 
was still able to collapse and fragment into a close binary protostar system,
while Price \& Bate (2007) found that magnetic pressure was more important for 
inhibiting binary fragmentation than either magnetic tension or braking. 
Machida et al. (2008) found that binary fragmentation could occur provided 
that the initial magnetic cloud core rotated fast enough, while Hennebelle 
\& Fromang (2008) and Hennebelle \& Teyssier (2008) found that magnetic 
clouds could fragment if an initial density perturbation was large enough. 
Joos et al. (2012) found that the initial direction of the magnetic field 
with respect to the rotation axis had an important effect on the outcome of
the collapse. 

 Hennebelle et al. (2011) pointed out that while magnetic fields
are important for cloud collapse, numerical convergence of MHD
codes is difficult to achieve, illustrating the importance of
careful testing (and eventually cross-comparison) of MHD code
results. This paper is directed toward performing a variety of new 
code tests and new code comparisons with the Enzo AMR MHD code, in order
to gain the necessary confidence to proceed with more realistic
models of the collapse of dense molecular cloud cores (e.g., Machida
et al. 2008; Duffin \& Pudritz 2009), and in particular to learn
if a previous series of pseudo-MHD models of magnetic field effects 
on cloud collapse (Boss 2009, 2007, 2005,..., 1997) is valid. Price \& Bate (2007) 
and B\"urzle et al. (2011) considered the collapse of spherical, magnetic cloud 
cores with initially uniform density and uniform magnetic field strengths,
the MHD version of the standard isothermal test case of Boss \& Bodenheimer
(1979). Here we perform a number of similar MHD calculations, with 
the goal of comparing the Enzo AMR MHD results with these previous 
results.	However, we begin by first testing the Enzo code on the Boss \&
Bodenheimer (1979) test case, and comparing the Enzo results with those 
obtained by a variety of previous studies (e.g., Burkert \& Bodenheimer 1993;
Myhill \& Boss 1993; Bate et al. 1995; Sigalotti 1997; Truelove et al. 1998; 
Kitsionis \& Whitworth 2002; Arreaga-Garcia et al. 2007). 

\section{Numerical Methods}

 Protostellar collapse calculations by necessity must cover variations
of many orders of magnitude in cloud density and length scales, from 
pc-sized molecular cloud cores to AU-sized first protostellar cores.
Early 3D hydrodynamic codes (e.g., Boss \& Bodenheimer 1979) used
numerical grids that contracted in the radial direction as the cloud
collapsed in order to maintain adequate spatial resolution.
More recently, AMR codes have been developed,
which allow numerical grid points to be inserted wherever they might be 
needed, and to be removed wherever they are not needed, for the most 
efficient calculation of problems with strongly varying length scales.
A number of such AMR codes exist at present and many are available for
public downloading (e.g., FLASH -- Fryxell et al. 2000; RIEMANN -- Balsara 
2001; RAMSES -- Teyssier 2002; NIRVANA -- Ziegler 2008, AstroBEAR --
Cunningham et al. 2009; Enzo -- Collins et al. 2010; CHARM -- Miniati 
\& Martin 2011; CRASH -- van der Holst et al. 2011; AZEuS -- Ramsey 
et al. 2012; PLUTO -- Mignone et al. 2012). 

 The Enzo2.0 code is used here, for the following reasons. Enzo performs
hydrodynamics (HD) using any one of three different algorithms: (1) the 
piecewise parabolic method (PPM) of Colella \& Woodward (1984), (2) the ZEUS 
method of Stone \& Norman (1992), or (3) a Runge-Kutta third-order-based MUSCL 
(``monotone upstream-centered schemes for conservation laws") algorithm 
based on the Godunov (1959) shock-handling HD method. Enzo is designed for 
handling strong shock fronts, which occur late in the protostellar collapse 
phase. This involves solving the Riemann problem (e.g., Godunov 1959)
for discontinuous solutions of a fluid quantity that should be conserved.
Enzo's default Riemann solver is that of Harten-Lax-van Leer (HLL), though
other solvers are available as well. We used MUSCL HD with the HLL
Riemann solver for all of the MHD models.

 Enzo 2.0 also performs ideal magnetohydrodynamics (MHD), using the 
Dedner et al. (2002) divergence cleaning method (Wang et al. 2008) in
concert with the MUSCL HD method noted above. Wang \& Abel (2009) have
shown that while the Dedner MHD method is unable to exactly conserve the
magnetic flux (i.e., keep the divergence of the field equal to zero), in
practice the non-zero field divergence that arises during a typical calculation
is not large enough to be dynamically important. Other versions of Enzo
employ the constrained transport MHD method for preserving the zero divergence 
of the magnetic field (Collins et al. 2010). 

 Enzo2.0 has been extensively tested on a wide variety of MHD test problems 
and found to perform well. Enzo solves for the gravitational fields within 
the cloud by using a Fast Fourier Transform method for the root grid and 
multigrid relaxation on subgrids. Enzo also includes the option of employing 
flux-limited diffusion approximation radiative transfer, needed for future 
comparisons with previous pseudo-MHD models (e.g., Boss 2009). 

 Enzo2.0 includes about 100 possible test cases, one of which is the Boss \& 
Bodenheimer (1979) 3D calculation for molecular cloud collapse and fragmentation, 
a convenient starting point for testing Enzo's ability to handle collapse 
calculations. The results of this test case with the Enzo2.0 code have not been
published to date. In fact, this test case will not run successfully when initialized
using the parameters provided in the Enzo2.0 code: either the hydrodynamic method 
to be used needs to be changed from the method involving the Dedner et al. (2002) 
MHD algorithm, or else the ``dual energy formalism'' switch must be disabled. 
In this paper we thus provide the first evaluation of Enzo2.0 on this test case,
and we also extend this basic collapse test case to consider the smaller initial 
amplitude density perturbation considered by Burkert \& Bodenheimer (1999), 
as well as higher spatial resolution. Given that Enzo is a Cartesian coordinate
code, it can conserve linear momentum, but not necessarily angular momentum,
unlike codes written in spherical or cylindrical coordinates (e.g., Boss \&
Bodenheimer 1979). Hence we also test here Enzo's ability to conserve angular
momentum.

 Most of the Enzo models presented here were initialized on a 3D Cartesian grid 
with 32 grid points in each direction, while one model used 64 grid points
in each direction. A maximum of 6 levels of refinement was permitted as 
needed during the collapses, with a factor of 2 refinement occurring for each 
level, so that the maximum possible effective grid resolution was  $2^6 = 64$ 
times higher than the initial resolution of $32^3$ or $64^3$, i.e., $2048^3$ 
or $4096^3$, respectively. Refinement was performed whenever necessary
to ensure that the Jeans length constraint (e.g., Truelove et al. 1997) was 
satisfied by a factor of 8; in one test case, this was increased to being 
satisfied by a factor of 16. After the models were run, a bug was
revealed in the Enzo2.0 code regarding the gas mean molecular weight,
which effectively reduced the gas temperature from 10 K to 3 K. [Greg Bryan 
informed us that this error in the parameter reading code of Enzo was 
corrected in the main Enzo repository by Elizabeth Tasker in June 2011, 
shortly after we downloaded our copy of the Enzo files.] Because the 
temperature is not a primary variable in Enzo, this bug had no direct effect
on the hydrodynamics, except that the Jeans length constraint that was
actually enforced, which is calculated using the temperature,
was then a factor of $\sim$ 2 times more conservative than
if the temperature had been correct. The effective Jeans length constraints
were then satisfied by factors of either $\sim$ 16 or $\sim$ 32 as a result.
By chance, this accidentally improved satisfaction of the Jeans length 
constraint compares favorably with the results of a recent study of AMR MHD 
(Federrath et al. 2011), which suggested that the Jeans length be resolved
by at least 30 cells in order to adequately resolve turbulence excited by 
gravitational collapse.

 Periodic boundary conditions were applied on
each face of the grid's cubic box. The maximum number of Green's functions
used to calculate the gravitational potential was 10. The time step used 
was 0.3 of the limiting Courant time step. The results were analyzed with 
the yt astrophysical analysis and visualization toolkit (Turk et al. 2011).

\section{Initial Conditions}

 The various models studied here all begin with variations on the 
initial conditions first studied by Boss \& Bodenheimer (1979) in what
has come to be known as the ``standard isothermal test case'' for 3D
hydrodynamics codes. The initial cloud is a uniform density ($\rho = 1.44 
\times 10^{-17}$ g cm$^{-3}$), uniform temperature ($T = 10$ K) sphere with 
radius $R = 3.2 \times 10^{16}$ cm, in solid body rotation around the $\hat z$ axis 
at an angular velocity $\Omega = 1.6 \times 10^{-12}$ rad s$^{-1}$. The cloud
is embedded within an active computational AMR grid with a linear dimension
of $3.2 \times 10^{17}$ cm, i.e., a 3D box with sides 10 times larger than
the cloud radius. The initial 
cloud has a mass $M = 1.0 M_\odot$, composed of molecular hydrogen (mean
molecular weight $\mu = 2.0$), an initial ratio of thermal to gravitational 
energy of $\alpha_i = 0.25$, and an initial ratio of rotational to gravitational 
energy of $\beta_i = 0.20$. 

 While a uniform density sphere is clearly a highly artificial representation
of a molecular cloud core, the Boss \& Bodenheimer (1979) initial conditions 
are nevertheless easy to reproduce for inter-code test comparisons, which
was the main motivation for their paper. These initial conditions were also
motivated by the earlier Black \& Bodenheimer (1976) axisymmetric,
two dimensional (2D) rotating cloud collapse calculations, which found that rings 
might form, which were themselves likely to fragment into multiple protostar 
systems in a fully 3D calculation. Boss \& Bodenheimer (1979) found that
when given an initial 50\% azimuthal ($cos(m \phi)$, with $m = 2$) density 
perturbation, the perturbation grew and fragmented the cloud into a binary 
protostar system. Isothermal thermodynamics with a temperature of $T = 10$ K 
was assumed throughout the collapse. 

 The Boss \& Bodenheimer (1979) standard isothermal test case, or
a slight variation with an initial 10\% azimuthal density perturbation, has
been often used to test 3D hydro codes (e.g., Burkert \& Bodenheimer 1993;
Myhill \& Boss 1993; Bate et al. 1995; Sigalotti 1997; Truelove et al. 1998; 
Kitsionis \& Whitworth 2002; Arreaga-Garcia et al. 2007). A nonisothermal test 
case for protostellar collapse also exists, calculated with two different 3D 
radiative hydrodynamics codes by Myhill \& Boss (1993), where compressional 
heating and radiative cooling (in either the Eddington or the diffusion 
approximation) are taken into account. In order to gain insight into how
MHD fragmentation might be affected by nonisothermal thermodynamics,
a set of MHD models has been calculated where the isothermal approximation
is replaced by the ``barotropic'' equation employed by, e.g., Price \& Bate 
(2007) and B\"urzle et al. (2011) in their 3D MHD models of the collapse of
uniform density spheres. The barotropic pressure $p$ is assumed to
depend on the gas density $\rho$ as $p = K \rho^\gamma$, where the
polytropic exponent $\gamma$ equals 1 for $\rho < 10^{-14}$ g cm$^{-3}$
and 7/5 for $\rho > 10^{-14}$ g cm$^{-3}$, mimicking the effects of 
nonisothermal thermodynamics. $K = c_s^2$, where $c_s$ is the isothermal 
sound speed (0.2 km s$^{-1}$), below the critical density of 
$\rho_c = 10^{-14}$ g cm$^{-3}$, while for $\rho > \rho_c$, 
$K = c_s^2 \rho_c^{-2/5}$, ensuring continuity of the pressure 
across the critical density. While a convenient approximation for
code inter-comparisons, the barotropic approximation does not
necessarily lead to the same outcome as a collapse calculated with
a full thermodynamical treatment, including Eddington approximation
radiative transfer (e.g., Boss et al. 2000), so some caution is
warranted when drawing conclusions from such models.

 The models with magnetic fields all began with the same initial conditions
as in the Boss \& Bodenheimer test case, with initial density perturbations of
either 10\% or 50\%, and with initially straight magnetic fields that were 
aligned with either the $\hat x$ or $\hat z$ axis (i.e., either perpendicular 
or parallel to the cloud's rotation axis, respectively). Table 2 lists the 
initial conditions and basic results for the MHD models with the isothermal
equation of state, while Tables 3 and 4 refer to the MHD models
with the barotropic equation of state and initial 50\% or 10\% density
perturbations, respectively. The results columns describe the midplane
($z = 0$) density configurations at the final time computed $t_f$, 
expressed in terms of the initial uniform density cloud free fall time 
$t_{ff} = (3 \pi / 32 G \rho)^{1/2} = 1.76 \times 10^4$ yr,
where $G$ is the gravitational constant and 
$\rho = 1.44 \times 10^{-17}$ g cm$^{-3}$.

\section{Isothermal HD Results}

 Three non-magnetic, isothermal models were computed, which were either
identical to Boss \& Bodenheimer (1979) [model BB79], or were a
variation on the standard isothermal test case: model BB79-10, where
the initial density perturbation was 10\% instead of 50\%, and model
BB79-10-HR, where the initial grid was 64$^3$ instead of 32$^3$ and
the Jeans length constraint was improved by another factor of 2.

 Figure 1 shows the evolution of model BB79. The large initial density
perturbation seen in Figure 1a collapses and fragments directly into
the binary protostar system seen in Figure 1d, at a maximum density
(nearly $\sim 10^{-11}$ g cm$^{-3}$) that is considerably higher than 
the maximum density for which the isothermal approximation is 
valid, i.e., beyond $\rho_c = 10^{-14}$ g cm$^{-3}$. This evolution
is quite similar to that previously obtained using a variety of 3D 
HD codes (e.g., Myhill \& Boss 1993; Burkert \& Bodenheimer 1993;
Bate, Bonnell \& Price 1995; Sigalotti 1997).

 A different evolution results when the initial density perturbation
is reduced from 50\% (model BB79) to 10\% (model BB79-10). As first
found by Burkert \& Bodenheimer (1993), the initial density
perturbation forms only a transient binary that merges into a thin
bar capable of fragmenting into multiple objects. Truelove et al. (1998)
found that the thin bar continued to collapse to very high densities
($\sim 10^{-10}$ g cm$^{-3}$) and formed two singular filaments. Similar 
thin bars were found by Kitsionis \& Whitworth (2002) and
by Arreaga-Garcia et al. (2007) with their smoothed particle
hydrodynamics (SPH) codes. 

 Figure 2 shows the evolution of model BB79-10. As a result of
the smaller initial density perturbation, the collapse and
fragmentation process proceeds slightly slower than in model BB79,
and the binary that forms (Figure 2b) is weakly defined compared
to model BB79. The binary fragments begin to merge together
(Figure 2c) into a bar-like structure, which then collapses
further to form a thin, intermediate filament (Figure 2d), similar 
to those previously found with other high-resolution HD codes.

 Finally, we consider model BB79-10-HR, with the spatial resolution
increased by a factor of two in all three directions compared
to model BB79-10, along with an improved Jeans length constraint. 
This model collapses in much the same
manner as model BB79-10, implying that the collapse is reasonably
well-resolved to the phase of formation of the thin filament.
Figure 3 shows that at this phase of evolution, both models are
quite similar, though the structure in model BB79-10-HR is
clearly better-defined than in model BB79-10, as expected.

 Enzo is designed to conserve mass and linear momentum, but not angular 
momentum. The degree to which mass and angular momentum are conserved
depends on the hydrodynamic method (HM) used, as well as on other parameters
in the Enzo input file. We have tested all three of the possible Enzo
hydrodynamics algorithms (see Numerical Methods section) for mass and 
angular momentum conservation on the Boss \& Bodenheimer (1979) test
case. The results are listed in Table 1. Model BB79 was run using the 
ZEUS method (Stone \& Norman 1992), model BB79-HM-0 with the PPM method 
(Colella \& Woodward 1984),and model BB79-HM-4 with the Runge-Kutta 
MUSCL Godunov (1959) method, denoted simply by MUSCL. Models BB79-10 and 
BB79-10-HR were also run with the MUSCL method. Table 1 shows
that the total mass $M_{tot}$ and total angular momentum $J_{tot}$
are reasonably well conserved with
all three methods, though the exact reason for even these relatively modest
losses is unclear, given that other 3D hydro codes are able to conserve
mass and angular momentum to machine precision (e.g., Boss 1997). The
most likely cause is the interpolation that is used to create new
sub-grids and remove old sub-grids. The models in Table 1 all used
Enzo's "second-order accurate" interpolation methods, which seemed
to perform the best on this problem. When Enzo's "first-order accurate"
interpolation was used, 4\% of the mass and 14\% of the angular
momentum was lost for BB79, compared to 0.7\% and 5\%, respectively,
for the "second-order accurate" interpolation method. With the use of this
latter interpolation method, Table 1 shows that the MUSCL hydrodynamics 
method is acceptable, and these methods were used for all of the MHD models
in the remainder of this paper.

 We conclude that Enzo is able to reliably reproduce previously
agreed-upon 3D HD results for the standard isothermal test case
and its variations, and that the initial spatial resolution
used ($32^3$) for these exploratory models is sufficient to
give an adequate indication of the likelihood of fragmentation
during the isothermal collapse regime.

\section{Isothermal MHD Results}

 We now turn to the effects of frozen-in magnetic fields on the 
collapse and fragmentation of initially uniform density, isothermal 
clouds. Tables 2 and 3 summarize the initial conditions and
outcomes of these models, which started from initial conditions
identical to either model BB79 (Table 2) or to model BB79-10 (Table 3), 
but with an initially straight magnetic field that was either parallel 
($B_z \ne 0$) or perpendicular ($B_x \ne 0$) to the initial rotation 
axis ($\hat z$). Given that model BB79 clearly fragmented into at
least a binary protostar system (Figure 1), while model BB79-10 formed
a thin bar, these two sets of models provide a convenient means
to assess the overall effects of frozen-in magnetic fields on
the collapse and fragmentation of dense cloud cores.

 Tables 2 and 3 also list the initial mass to flux ratios ($M/\Phi$)
for the models, given in units of the nominal critical ratio for 
stability, based on the values given by Price \& Bate (2007).
Thus models with $M/\Phi > 1$ are magnetically supercritical and 
nominally unstable to gravitational collapse, while models 
with $M/\Phi < 1$ are magnetically subcritical and nominally
stable to gravitational collapse. 

\subsection{BB79 MHD Results}

 Three distinct outcomes characterize the magnetic cloud models: (1) contraction 
over a few orders of magnitude increase in density, without sustained collapse,
resulting in an un-fragmented, though considerably denser, magnetic cloud core,
(2) sustained collapse to form a single protostar with significant spiral arms, 
and (3) collapse and fragmentation into binary or multiple protostar systems, 
with multiple spiral arms. In the language of magnetic field theory (e.g.,
Crutcher 2012), evidently outcome (1) involves a magnetically subcritical
cloud, while outcomes (2) and (3) involve magnetically supercritical clouds.

 Figure 4 shows the evolution of model mag-z-300, which evolved the
closest to model BB79, given that this model had the lowest initial
magnetic field strength of the $B_z \ne 0$ models, coupled with the
fact that the $B_z \ne 0$ models allow for the gas to collapse unimpeded
along the magnetic field lines onto the $z = 0$ midplane, 
which is also favored by rotation about the $\hat z$ axis. Compared
to Figure 1, Figure 4 shows that the magnetic field prevents the initial
density perturbations from growing as rapidly as in BB79, allowing
the perturbations to fall inward closer together (Figure 4b) before forming
two thin filaments (Figure 4c) which ultimately fragment into a binary or
multiple protostellar system (Figure 4d), accompanied by a large number
of spiral arms and moderate density clumps. While the magnetic field is
able to slow the growth of the density perturbation, forcing it into
an intermediate thin filament similar to that produced in BB79-10,
in the end the frozen-in magnetic field is not strong enough to
prevent the cloud from fragmenting. By the time this fragmentation
occurs, however, the maximum density ($> 10^{-12}$ g cm$^{-3}$) 
exceeds that where the isothermal approximation remains valid, so
this result should not be taken as fully realistic.

 Model mag-z-300 can be used to test the degree to which Enzo conserves
magnetic flux, as should be the case for ideal hydrodynamics. This is
accomplished by monitoring the divergence of the magnetic field ("DivB"),
which starts equal to zero everywhere in the cloud, and should remain
so throughout the evolution. Requiring a zero divergence of magnetic field is
equivalent to stating that there are no magnetic charges, in contrast
to electric charges. In practice, the Dedner el al. (200-2) MHD method 
does not keep DivB = 0, but Enzo k.eps DivB small enough that the errors
that are introduced as a result are not severe (Wang \& Abel 2009). We
have tested this assertion for model mag-z-300 by monitoring DivB throughout
the evolution depicted in Figure 4. For the initial cloud (Figure 4a),
DivB = 0 everywhere, as expected. As the evolution proceeds, DivB begins
to grow, first at the edge of the cloud, and then later at the boundaries
of new sub-grids as the collapse proceeds (Figure 4b). By the time that
fragmentation has occurred (Figure 4c,d), DivB peaks at the locations
of the density and magnetic field maxima. The dynamical importance of 
non-zero DivB can be assessed by calculating the ratio of DivB to the ratio 
of the magnetic field strength divided by the local grid spacing, as was
done by Wang \& Abel (2009), i.e., $\nabla \cdot B \Delta x / B$.
This critical ratio should be less than one if the MHD solution is
to be reasonably accurate (Wang \& Abel 2009). When evaluated for
the highest peaks of DivB for model mag-z-300 (Figure 4d), 
$\nabla \cdot B \Delta x / B \sim < 0.1$, showing that, while not perfect, 
Enzo2.0 is able to conserve the magnetic flux reasonably well. 

 Figure 5 shows the results for two models that illustrate the other 
two possible outcomes noted above: model mag-z-1600, which collapses
to form a single protostar with spiral arms, and model mag-z-1800,
which contracts but does not collapse. These two models demonstrate
the dramatic difference in outcome produced by a relatively
small change in the initial magnetic field strength: evidently
these two models span a critical value for dynamic collapse versus
minor contraction, for the assumed initial conditions.

 Figure 6 shows the evolution of model mag-x-300, identical to model
mag-z-300 except for the initial orientation of the magnetic
field. While both models fragment, the evolutions are significantly
different, as can be seen by comparing Figures 4 and 6. Because gas can 
flow along the field lines aligned initially with the $\hat x$ axis in model
mag-x-300, Figure 6b shows that the initial density perturbations
(which are also aligned initially with the $\hat x$ axis) grow 
preferentially along the $\hat x$ axis direction, compared
to Figure 4b for model mag-z-300, where the growth is more in the
radial direction. This leads the formation of two off-set, nearly
parallel filaments in the former model (Figure 6c), compared to
two filaments with a linear alignment in the latter (Figure 4c). The
subsequent fragmentation of the filaments in either case leads
to a binary or multiple protostellar system (Figures 4d and 6d),
though with distinct overall differences at this early phase of
evolution.

 Model mag-x-300 has also been tested for magnetic flux conservation,
as was done for mag-z-300, and the ratio $\nabla \cdot B \Delta x / B$
was found to be no worse than 0.1, again implying a reasonably accurate
MHD solution.

 Figure 7 compares the results for models mag-x-800 and mag-x-1000:
the former collapses to form a single protostar with multiple
spiral arms, while the latter contracts but does not collapse.
The center of the cloud in model mag-x-1000 does shift along
the $\hat x$ axis during its lengthy (over 20 $t_{ff}$) evolution,
which is permitted because of the assumption of periodic boundary
conditions for these models. While the initial cloud is axisymmetric,
non-axisymmetry arises during the evolutions as a result of growth
of numerical round-off error, accounting for the asymmetric appearance
of many of the models that were evolved for long times. The motion 
of the center of mass of the cloud for model mag-x-1000 is along the 
direction of the initial ($\hat x$) magnetic field lines, so this 
overall motion is consistent with movement along field lines. 
A considerably smaller lateral shift is evident for model mag-z-1800
(Figure 5b), where the initial magnetic field retards such motions.

 Flow down magnetic field lines compared to flow across field lines 
leads to differing effects, as can be seen in Figure 8, which compares 
the vertical density distributions for two models which both 
collapsed to form single protostars, models mag-z-1600 (Figure 8a)
and mag-x-800 (Figure 8b). Figure 8a shows that the mag-z-1600 cloud
collapsed along the $\hat z$ axis, to form a fairly symmetrical
distribution about the $z = 0$ plane, as both the initial rotation
and magnetic fields permit collapse into this plane. Figure 8b
shows that model mag-x-1600 did collapse primarily into the
$z = 0$ plane, as permitted by the initial rotation axis, but
the vertical distribution is considerably less symmetrical about
the $z = 0$ plane, as a result of the resistance of the initial
magnetic field to collapse along the $\hat z$ axis.

 Table 2 shows that for initial 50\% density perturbations,
the initial magnetic field direction compared to the rotation
axis has a profound effect on the outcome of the collapse.
We can define several critical initial magnetic field strength
values for these initial conditions: $B_{coll}$ is the maximum
field strength permitting sustained collapse, while $B_{frag}$ 
is the maximum field strength permitting collapse leading to
fragmentation. These two critical values differ significantly
depending on the initial field orientation: for fields aligned
with the rotation axis, $B_{coll} \sim 1700 \mu$G and
$B_{frag} \sim 1550 \mu$G, whereas for fields initially aligned
perpendicular to the rotation axis, $B_{coll} \sim 900 \mu$G and
$B_{frag} \sim 500 \mu$G. The fact that both critical values
are considerably lower in the latter case than in the former
can be qualitatively understood in part as being the result of the
ability of the dense cloud core to collapse relatively unimpeded
to the $z = 0$ midplane (unopposed by rotation or magnetic fields)
in the former case, compared to the inhibition to collapse to 
the $z = 0$ midplane when the field is initially perpendicular 
to the rotation axis in the latter case. The extent of angular
momentum loss suffered due to magnetic braking in these clouds
(Table 2) is clearly also a factor, as for both field orientations,
once the total angular momentum loss reaches $\sim$ 50\%, 
fragmentation into a binary or multiple system is stymied.
That this angular momentum loss occurs for smaller initial
field strengths than for the perpendicular field alignment implies
that such perpendicular fields can be more efficient at magnetic 
braking, i.e., stretching of the initial $B_x$ field lines transports
angular momentum outward more efficiently than does twisting
of the initial $B_z$ field lines.

 These results provide a
quantitative estimate of how the initial magnetic field value
can lead to significantly different outcomes for dense cloud 
core evolutions. For comparison, observational Zeeman effect
estimates of the line-of-sight component of the magnetic field  
in dense cloud cores with number densities appropriate for these 
(BB79) models ($\sim 10^7$ cm$^{-3}$) are on the order of
$B_{los} \sim 1000 \mu$G (see Figure 6 in Crutcher 2012), making these
critical field strengths of more than theoretical importance 
for real molecular cloud cores.

 It is also interesting to compare these results with the magnetic
field strength that separates formally magnetically subcritical
clouds from magnetically supercritical clouds (e.g., Nakano \&
Nakamura; Crutcher 2012). For a uniform density cloud threaded
by a uniform magnetic field, the critical field strength is
$B_c \approx 2 G^{-1/2} M / R^2$.
For the BB79 cloud, $B_c \approx 1000 \mu$G, which falls between 
the maximum initial field strengths for collapse (Table 2) of 
$B_{coll} \sim 1700 \mu$G for fields aligned with the rotation 
axis, and $B_{coll} \sim 900 \mu$G for perpendicular fields.
Note that based on the Price \& Bate (2007) values for the critical mass
to flux ratio, this critical field strength would be $\sim 800 \mu$G,
just slightly lower. The numerical models thus straddle the expectations 
based on the analytic criterion for non-rotating clouds, where the field
orientation can have no effect, as well as based on the initial
magnetic flux ratio values given in the Tables. 

\subsection{BB79-10 MHD Results}

 We now turn to the models listed in Table 3, which are identical
to those just discussed, except for having a smaller (10\%) initial 
density perturbation, as in the BB79-10 models. The same three 
general types of outcomes are found for these models, as noted above 
and as listed in Table 3. These models were restricted to having
the magnetic field aligned with the rotation axis.

 The key result is that the critical magnetic field strength values 
(Table 3) are significantly different from those obtained 
with the initial 50\% density perturbations (Table 2).
The critical values of $B_{coll} \sim 1700 \mu$G and
$B_{frag} \sim 1550 \mu$G for 50\% perturbations decrease to
$B_{coll} \sim 1350 \mu$G and $B_{frag} \sim 1050 \mu$G for 10\%
perturbations, reflecting the fact that with smaller initial 
density perturbations, the magnetic fields must be substantially
smaller in order to still achieve collapse and fragmentation,
though still in the range of observed dense cloud cores (e.g.,
Crutcher 2012).
 
\section{Barotropic MHD Results}

 Collapsing clouds that are assumed to remain isothermal are considerably
more likely to undergo fragmentation than clouds where compressional
heating begins to raise the cloud's temperature above $\sim$ 10 K at
densities above $\sim 10^{-14}$ g cm$^{-3}$ (e.g., Boss 1986). Accordingly,
we now turn to the results for the MHD models with the same barotropic 
equation employed by Price \& Bate (2007) and B\"urzle et al. (2011) in 
their 3D MHD models of the collapse of uniform density spheres. These
models should afford a more realistic assessment of the prospects for
magnetic cloud collapse and fragmentation than the isothermal models,
at least for clouds which collapse into the nonisothermal regime before
fragmenting. Table 4 shows the results for the barotropic MHD models.
All of the barotropic models started with $B_z < B_c \sim 1000 \mu$G,
i.e., initially magnetically supercritical, so all of these models collapsed.
The goal of this series was to determine $B_{frag}$, rather than $B_{coll}$,
and to compare $B_{frag}$ with the previous results.

 Figure 9 shows the evolution of model poly-z-20, which collapsed to
form an intermediate central object with spiral arms (Figure 9b),
which continue to grow as further mass collapsed to the midplane,
leading to fragment formation near the ends of the two-armed spiral
(Figure 9c). One of these two new fragments merged with the central
object, while the second object was still in orbit at the final time
shown (Figure 9d). 

 Figure 10 shows the evolution of model poly-z-81, which collapsed to
form an intermediate central bar (Figure 10b). The bar evolved into
a central object surrounded by spiral arms (Figure 10c), but in this
case, the accretion of further mass into the midplane did not result
in the formation of additional clumps: at the same final time as
model poly-z-20, a single central object with spiral arms is
evident (Figure 10d).

 Models poly-z-20 and poly-z-81 represent the two types of
outcomes for this set of models (Table 4): collapse in all cases,
but sustained fragmentation only for the models with 
$B_z < B_{frag} \sim 70 \mu$G. This value for $B_{frag}$ is 
considerably lower than those obtained for the previous sets
of isothermal MHD models, a clear indication of the added difficulty
for fragmentation when the cloud is allowed to heat up upon becoming
optically thick at infrared wavelengths at $\rho \sim 10^{-14}$ g cm$^{-3}$.

 This critical value of $B_{frag} \sim 70 \mu$G is also somewhat
lower than the critical value found in two previous 3D MHD studies using
the barotropic pressure relationship with similar initial conditions.
Price \& Bate (2007) found $B_{frag} \sim 180 \mu$G for initial fields
aligned with the rotation, as is the case here (Table 4), and
$B_{frag} \sim 95 \mu$G when the field was perpendicular to the rotation
axis. B\"urzle et al. (2011) found $B_{frag} \sim 300 \mu$G for initial fields
aligned with the rotation. The reason for the differences in
$B_{frag}$ between these two previous studies is unclear, as the
later study was intended to duplicate the earlier study, and both
studies used SPH MHD codes with the same barotropic relation, starting
from identical initial conditions. However, B\"urzle et al. (2011)
suggested that one reason for the differences could be due to the
different MHD techniques employed, resulting in different degrees
of ``magnetic cushioning'', which could aid fragmentation.
In the present models, magnetic cushioning does not appear to play as
large a role, as the maximum absolute values of $B_x$ and $B_y$ are about 
three times as large as $B_z$ in the midplanes of both models poly-z-60
and poly-z-81 at their final times, when the former had fragmented, while
the latter had not (Figure 10d). These maximum field strengths are all significantly
higher (factors of 2 to 3) in model poly-z-81 than in poly-z-60 at these final times, 
in spite of their initial field strengths being similar. These final field strengths
presumably then are the main reason for the different outcomes in models poly-z-60
and poly-z-81.

 Besides the different MHD techniques employed, the differences between 
these two previous SPH studies and the models shown in Table 4 are likely
to be due in part to the relatively modest spatial resolution employed in
the present calculations, compared to the use of $\sim 3 \times 10^6$
particles in the SPH calculations, combined with the use of sink cells
in the SPH models. These sink cells are inserted in both sets of 
SPH models once the density exceeds $10^{-10}$ g cm$^{-3}$.
While it is not clear from the figures presented in the two SPH
papers, the sink cells appear to be inserted in both calculations
at about the phase of evolution seen in Figure 10b, for models with
initial $B_z = 81 \mu$G for all three codes. The maximum midplane
density for model poly-z-81 seen in Figure 10b is only 
$\sim 10^{-13}$ g cm$^{-3}$, much less than the sink cell critical
density, so clearly the SPH codes are resolving much higher densities
during these collapses than are being resolved with Enzo with
the moderate resolution employed. This appears to be the main
reason for the lower value of $B_{frag}$ for the poly-z models,
as other differences with the SPH models (e.g., small differences
in the initial cloud density, radius, temperature, and rotation rate) 
appear to be less important. At any rate, the differences in the
$B_{frag}$ values between the poly-z models and the SPH results
are comparable to the differences between the two SPH codes themselves;
overall, all three codes are in good agreement about the outcome
of these 3D MHD collapse and fragmentation models.

\section{Conclusions}

 We have tested the Enzo AMR code on a variety of 3D HD and MHD
test cases for protostellar collapse, and shown that the Enzo code 
performs reliably, even with only moderate spatial resolution, and 
yields results in good agreement with previous studies. We have tested
the performance of various hydrodynamical methods on the Boss \& 
Bodenheimer (1979) standard isothermal test case, and determined
the degree to which the various methods conserve mass and angular
momentum. The MHD models
have led to estimates of the critical initial magnetic field strengths 
($B_{coll}$) that separate contraction from collapse (i.e., magnetically
subcritical versus supercritical clouds), as well as the strengths
($B_{frag}$) that separate collapse to form a single protostar
from those that lead to fragmentation into a binary or multiple
protostar system. The different fragmentation outcomes depend in part 
on the amount of angular momentum lost by magnetic braking during the
evolutions, with the isothermal MHD models implying that magnetic
braking is more efficient for perpendicular than for aligned fields. The 
numerically determined values of $B_{coll} \sim 1700 \mu$G for initial 
fields aligned with the rotation axis and $B_{coll} \sim 900 \mu$G 
for perpendicular fields compare well with observational estimates 
of $B_{los} \sim 1000 \mu$G for the densest molecular cloud cores,
implying that factors such as the relative orientation of the
magnetic field and the rotation axis could have an important
effect on cloud collapse.
 
 The initially uniform density, initially uniform rotation 
(non-turbulent) spheres considered here are convenient
for theoretical comparisons, but are not particularly similar to
observed molecular cloud cores, which are typically centrally-condensed 
(e.g., Ward-Thompson, Motte, \& Andr\'e 1999; Schnee et al. 2010; 
Butler \& Tan 2012), highly non-spherical, and turbulent. 
In particular, prolate and oblate shapes have been inferred from 
observations of suspected pre-collapse molecular cloud cores (e.g., 
Jones, Basu, \& Dubinski 2001; Curry \& Stahler 2001; Cai \& Taam 2010). 

 Armed with the experience gained from the present set of MHD models,
our future Enzo AMR MHD models will consider the collapse of more realistic, 
centrally-condensed, spheroidal molecular cloud cores, such as the 
initially prolate and oblate cloud cores considered by Boss (2009), 
using a pseudo-MHD code to model magnetic field effects and ambipolar
diffusion in a variety of clouds (e.g., Boss 2007, 2005, and references
therein). We hope to learn if these pseudo-MHD calculations produced
results similar to what can be obtained with a true MHD code like Enzo.

\acknowledgments

 We thank Tom Abel, Greg Bryan, Nathan Goldbaum, Craig Lage, Brian 
O'Shea, Britton Smith, and Matt Turk for advice about the Enzo and yt
codes, and the referee, Sven Van Loo, for a large number of great 
suggestions for improving
the manuscript. The computations were performed using the Enzo code 
developed by the Laboratory for Computational Astrophysics at the 
University of California San Diego (http://lca.ucsd.edu). The 
calculations were performed on the Carnegie Xenia Cluster, the purchase 
of which was partially supported by the National Science Foundation (NSF) 
under grant MRI-9976645. This work was partially supported by NSF grant
AST-1006305. We also thank Michael Acierno and Ben Pandit for their 
invaluable assistance with the Xenia cluster.

\clearpage
\begin{deluxetable}{cccc}
\tablecaption{Percentage change in the total mass ($M_{tot}$) and total
angular momentum ($J_{tot}$) during the evolutions of the isothermal HD 
models, showing the dependence on the hydrodynamic method chosen in
several variations of model BB79.
\label{tbl-1}}
\tablehead{\colhead{\quad \quad model \quad \quad} & 
\colhead{\quad \quad hydro \quad \quad} &
\colhead{\quad \quad $\Delta M_{tot}$ \quad \quad} &
\colhead{\quad \quad $\Delta J_{tot}$ \quad \quad} }
\startdata

BB79        &  ZEUS   & -0.7\%  & -5\%    \\

BB79-HM-0   &  PPM    & -0.5\%  & -2\%     \\

BB79-HM-4   &  MUSCL  & -0.7\%  & -4\%     \\

BB79-10     &  MUSCL  & -0.5\%  & -2\%     \\

BB79-10-HR  &  MUSCL  & -0.1\%  & -0.2\%    \\

\enddata
\end{deluxetable}
\clearpage

\clearpage
\begin{deluxetable}{ccccccc}
\tablecaption{Initial magnetic field strengths (in microGauss),
initial mass to flux ratios $M/\Phi$ (in units of the critical ratio), 
and the results (at the final time $t_f$, in free fall units) for 
the isothermal MHD models with initial 50\% density perturbations,
as in the non-magnetic model BB79.
\label{tbl-2}}
\tablehead{\colhead{model } & 
\colhead{$B_x$ ($\mu$G)} &
\colhead{$B_z$ ($\mu$G)} &
\colhead{$M/\Phi$} &
\colhead{$t_f/t_{ff}$ \quad } & 
\colhead{\quad $\Delta J_{tot}$ \quad} &
\colhead{\quad result} }
\startdata

mag-z-300  &  0  &  300  &  2.7   & 1.8 &  -5.0\%  & collapses/fragments  \\

mag-z-600  &  0  &  600  &  1.3   & 2.4 &  +3.1\%  & collapses/fragments  \\

mag-z-1000 &  0  & 1000  &  0.80  & 1.7 &  -5.9\%  & collapses/fragments  \\

mag-z-1200 &  0  & 1200  &  0.67  & 2.4 &  -8.6\%  & collapses/fragments  \\

mag-z-1300 &  0  & 1300  &  0.62  & 2.2 &  -9.1\%  & collapses/fragments  \\

mag-z-1400 &  0  & 1400  &  0.57  & 2.6 &  -13.\%   & collapses/fragments  \\

mag-z-1500 &  0  & 1500  &  0.53  & 6.0 &  -39.\%   & collapses/fragments  \\

mag-z-1600 &  0  & 1600  &  0.50  & 7.7 &  -49.\%   & collapses/single     \\

mag-z-1800 &  0  & 1800  &  0.44  & 9.2 &  -57.\%   & little collapse/single \\             

mag-z-2000 &  0  & 2000  &  0.40  & 11.5 & -63.\%   & little collapse/single \\

mag-x-300  & 300 &    0  &   2.7  & 5.8 &  -0.90\%  & collapses/fragments  \\

mag-x-400  & 400 &    0  &   2.0  & 3.2 &  -19.\%   & collapses/fragments  \\

mag-x-600  & 600 &    0  &   1.3  & 13.4 & -69.\%   & collapses/single     \\

mag-x-700  & 700 &    0  &   1.1  & 15.6 & -77.\%   & collapses/single     \\

mag-x-800  & 800 &    0  &   1.0  & 8.8 &  -67.\%   & collapses/single     \\

mag-x-1000 & 1000&    0  &  0.80  & 20.7 & -83.\%   & little collapse/single \\

mag-x-1200 & 1200&    0  &  0.67  & 44.0 & -82.\%   & little collapse/single \\

\enddata
\end{deluxetable}
\clearpage

\clearpage
\begin{deluxetable}{cccccc}
\tablecaption{Initial magnetic field strengths $B_z$ (in microGauss),
initial mass to flux ratios $M/\Phi$ (in units of the critical ratio), 
and the results (at the final time $t_f$, in free fall units) for 
the isothermal MHD models with initial 10\% density perturbations,
as in the non-magnetic model BB79-10.
\label{tbl-3}}
\tablehead{\colhead{\quad model } & 
\colhead{\quad $B_z$ ($\mu$G) \quad} &
\colhead{\quad $M/\Phi$ \quad} &
\colhead{\quad $t_f/t_{ff}$ \quad} & 
\colhead{\quad \quad $\Delta J_{tot}$ \quad \quad} &
\colhead{\quad \quad result \quad } }
\startdata

mag-z-400-10    &  400  &  2.0   &  2.2 &  +9.7\%  &  collapses/fragments  \\ 

mag-z-800-10    &  800  &  1.0   &  2.0 &  -1.2\%  &  collapses/fragments  \\

mag-z-900-10    &  900  &  0.89  &  3.5 &  -11.\%  &  collapses/fragments  \\

mag-z-1000-10   & 1000  &  0.80  &  7.3 &  -30.\%  &  collapses/fragments  \\

mag-z-1100-10   & 1100  &  0.73  &  9.5 &  -38.\%  &  collapses/single  \\

mag-z-1200-10   & 1200  &  0.67  &  3.5 &  -19.\%  &  collapses/single  \\
 
mag-z-1300-10   & 1300  &  0.62  & 14.0 &  -64.\%  &  collapses/single  \\

mag-z-1400-10   & 1400  &  0.57  &  6.1 &  -40.\%  &  little collapse/single  \\

mag-z-1600-10   & 1600  &  0.50  & 12.8 &  -61.\%  &  little collapse/single  \\

mag-z-1800-10   & 1800  &  0.44  & 15.8 &  -66.\%  &  little collapse/single \\

\enddata
\end{deluxetable}
\clearpage

\clearpage
\begin{deluxetable}{cccccc}
\tablecaption{Initial magnetic field strength $B_z$ (in microGauss),
initial mass to flux ratios $M/\Phi$ (in units of the critical ratio), 
and the results (at the final time $t_f$, in free fall units) for 
the barotropic MHD models, models that are otherwise identical
to the non-magnetic, isothermal model BB79-10.
\label{tbl-4}}
\tablehead{\colhead{\quad model } & 
\colhead{\quad $B_z$ ($\mu$G) \quad} &
\colhead{\quad $M/\Phi$ \quad} &
\colhead{\quad $t_f/t_{ff}$ \quad } & 
\colhead{\quad \quad $\Delta J_{tot}$ \quad \quad} &
\colhead{\quad \quad result \quad } }
\startdata

poly-z-407  &  407.0  &  2.0 &  1.85   &   -1.8\%   &   collapses/single   \\

poly-z-203  &  203.0  &  4.0 &  2.09   &   -1.1\%   &   collapses/single   \\  

poly-z-163  &  163.0  &  5.0 &  1.97   &   -1.9\%   &   collapses/single   \\

poly-z-109  &  109.0  &  7.5 &  2.20   &   -3.8\%   &   collapses/single   \\

poly-z-81   &   81.3  &  10. &  2.20   &   -4.9\%   &   collapses/single   \\

poly-z-60   &   60.0  &  15. &  2.20   &   -4.1\%   &   collapses/fragments  \\

poly-z-41   &   41.0  &  20. &  2.20   &   -3.1\%   &   collapses/fragments   \\

poly-z-20   &   20.0  &  41. &  2.20   &   -3.5\%   &   collapses/fragments   \\

poly-z-10   &   10.0  &  82. &  2.20   &   -2.4\%   &   collapses/fragments  \\

poly-z-5    &    5.0  &  164.&  2.20   &   -3.2\%   &   collapses/fragments   \\

poly-z-0    &    0.0  & $\infty$ & 2.20 &  -1.7\%   &  collapses/fragments   \\          

\enddata
\end{deluxetable}
\clearpage

\begin{figure}
\vspace{-2.0in}
\plotone{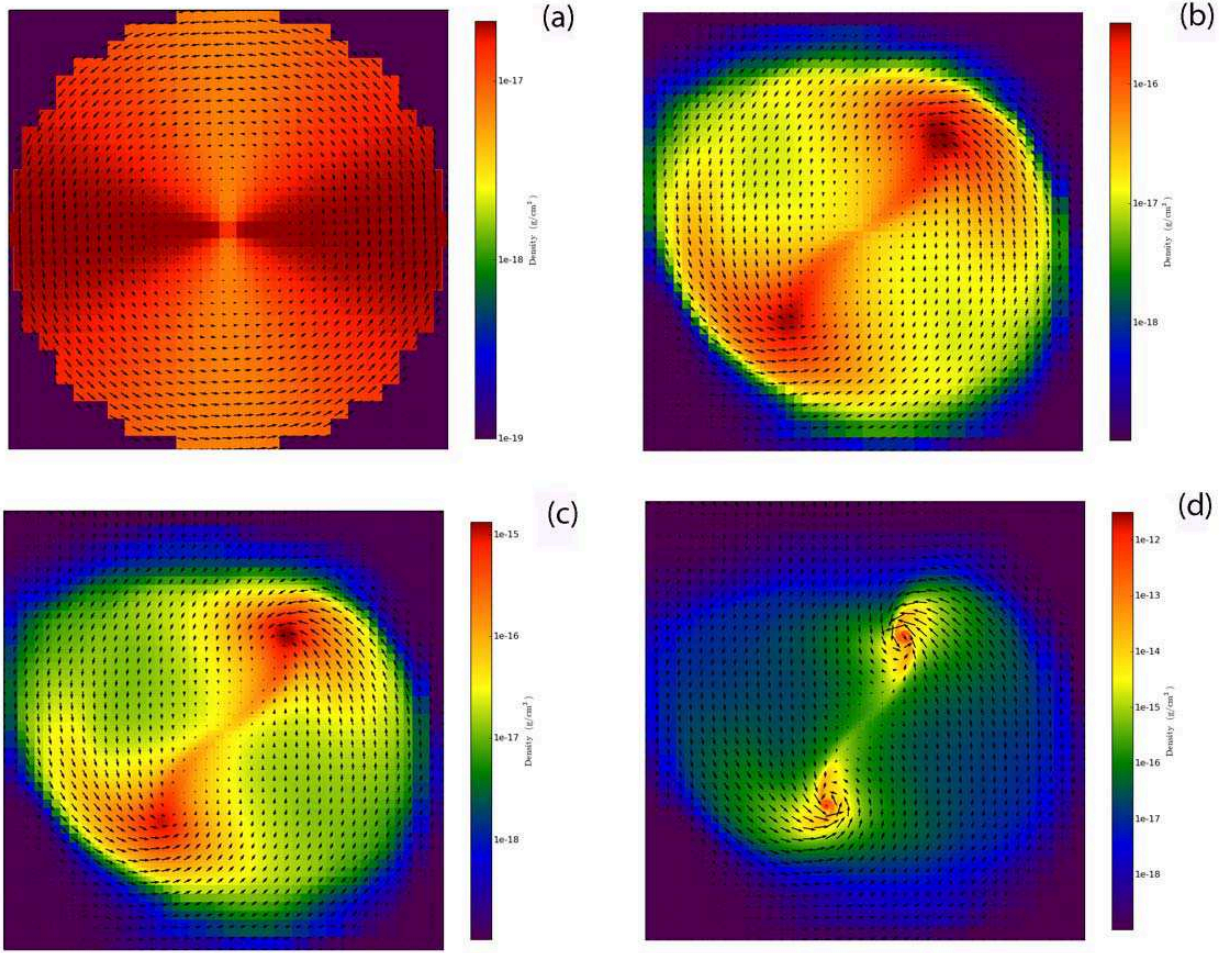}
\caption{Evolution of the midplane ($z = 0$) density distribution
and velocity vectors for the isothermal, non-magnetic model BB79,
shown at times: a - 0.0, b - 0.871 $t_{ff}$, c - 0.950 $t_{ff}$, 
and d - 1.100 $t_{ff}$. Maximum velocities are $\sim$ 0.5 km s$^{-1}$.
Region shown is $6.4 \times 10^{16}$ cm across in each case.}
\end{figure}

\clearpage

\begin{figure}
\vspace{-2.0in}
\plotone{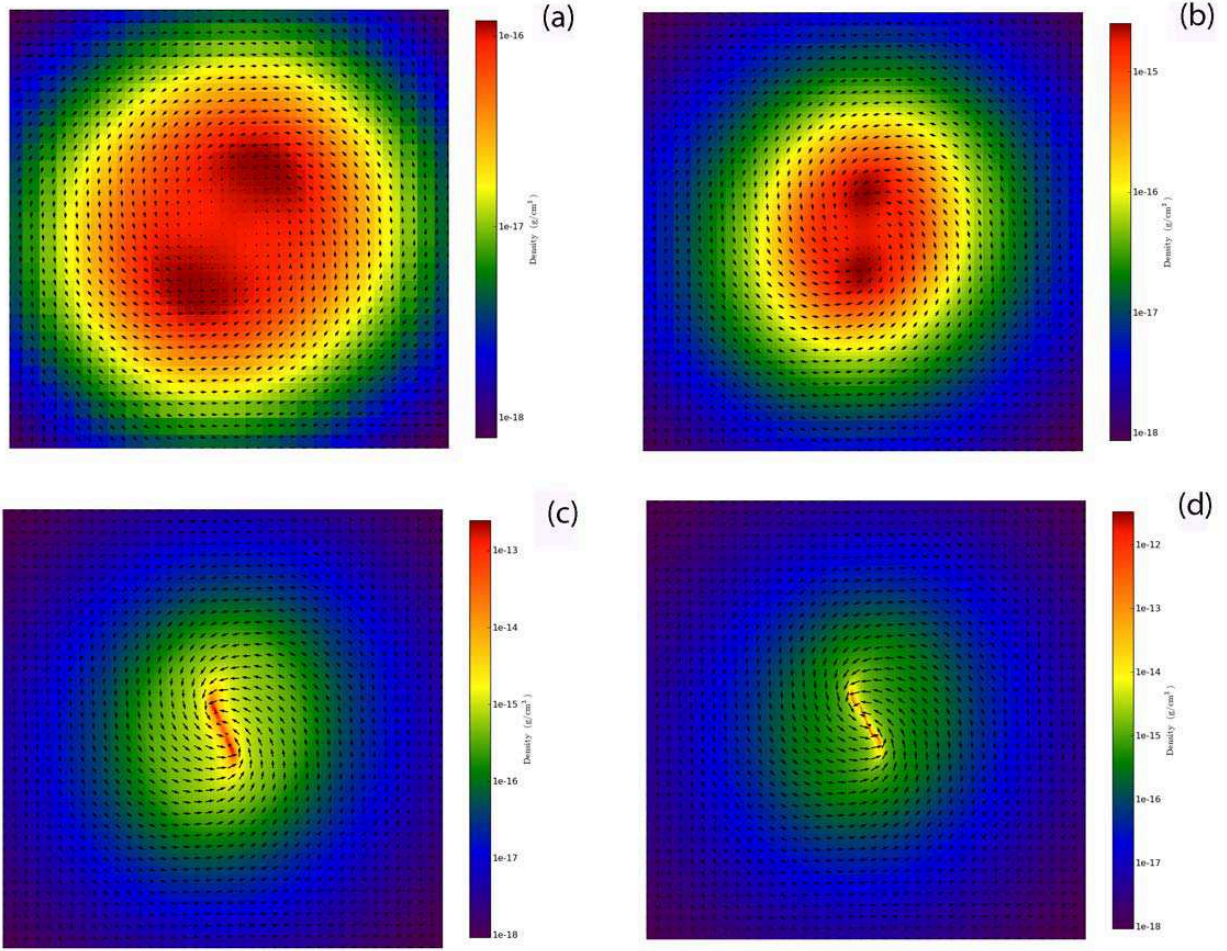}
\caption{Evolution of the midplane density distribution
and velocity vectors for the isothermal, non-magnetic model BB79-10,
shown at times: a - 0.990 $t_{ff}$, b - 1.230 $t_{ff}$, c - 1.364 $t_{ff}$, 
and d - 1.386 $t_{ff}$, plotted as in Figure 1.}
\end{figure}

\clearpage

\begin{figure}
\vspace{-2.0in}
\plotone{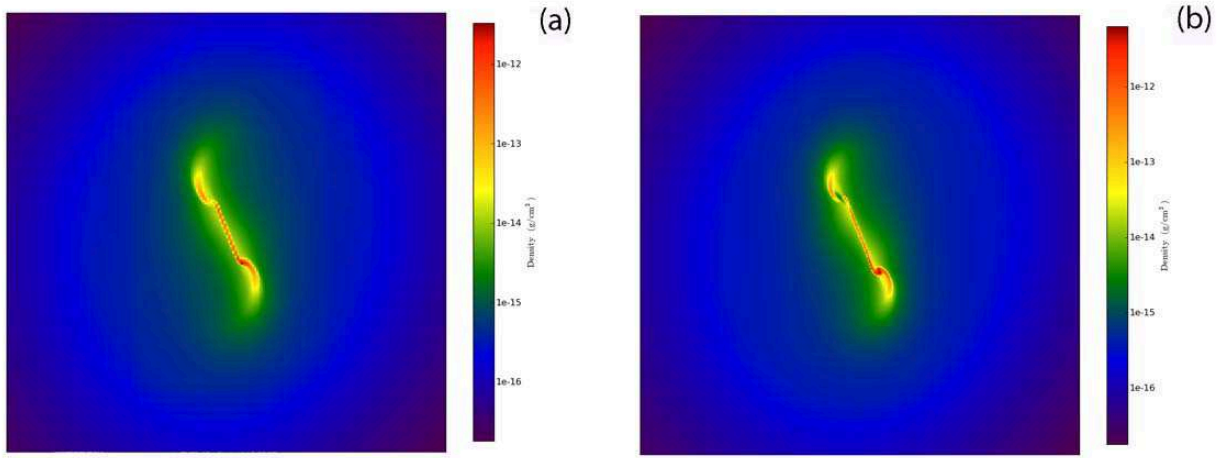}
\caption{Midplane density distributions for model BB79-10 (a) and for
model BB79-10-HR (b), both at a time of 1.386 $t_{ff}$. Region shown is 
$3.2 \times 10^{16}$ cm across in both cases.}
\end{figure}

\clearpage

\begin{figure}
\vspace{-2.0in}
\plotone{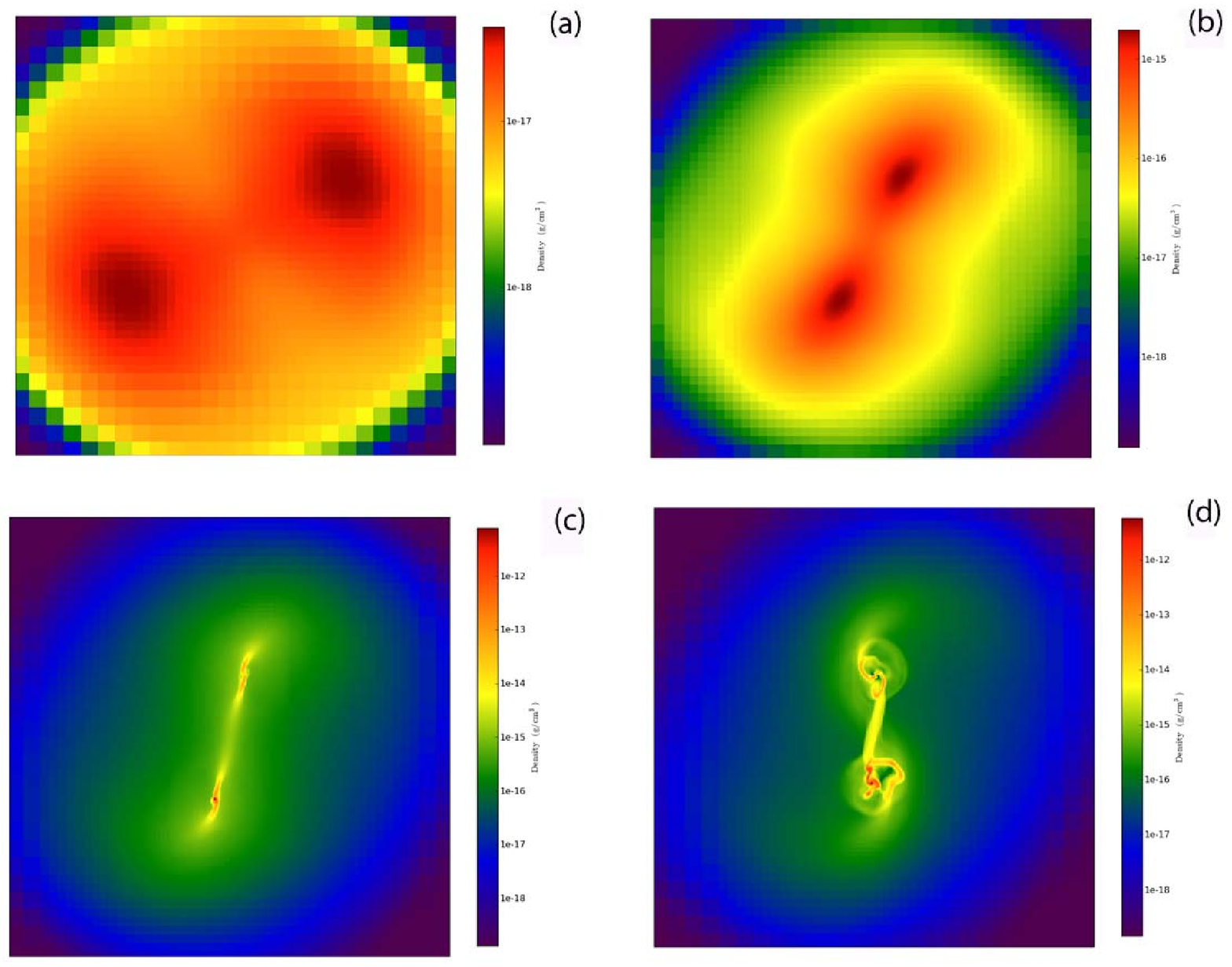}
\caption{Evolution of the midplane density distribution for the isothermal, 
magnetic model mag-z-300 (a magnetic version of model BB79),
shown at times: a - 0.530 $t_{ff}$, b - 1.109 $t_{ff}$, c - 1.324 $t_{ff}$, 
and d - 1.577 $t_{ff}$, plotted as in Figure 1.}
\end{figure}

\clearpage

\begin{figure}
\vspace{-2.0in}
\plotone{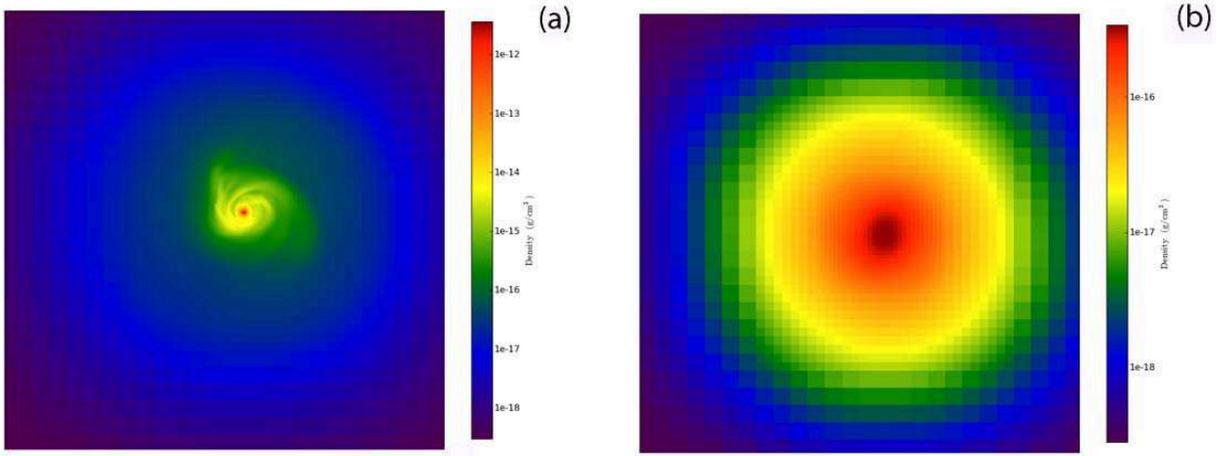}
\caption{Midplane density distributions for models mag-z-1600 (a) and 
mag-z-1800 (b), at times of 7.742 and 9.350 $t_{ff}$, respectively, plotted
as in Figure 1.}
\end{figure}

\clearpage

\begin{figure}
\vspace{-2.0in}
\plotone{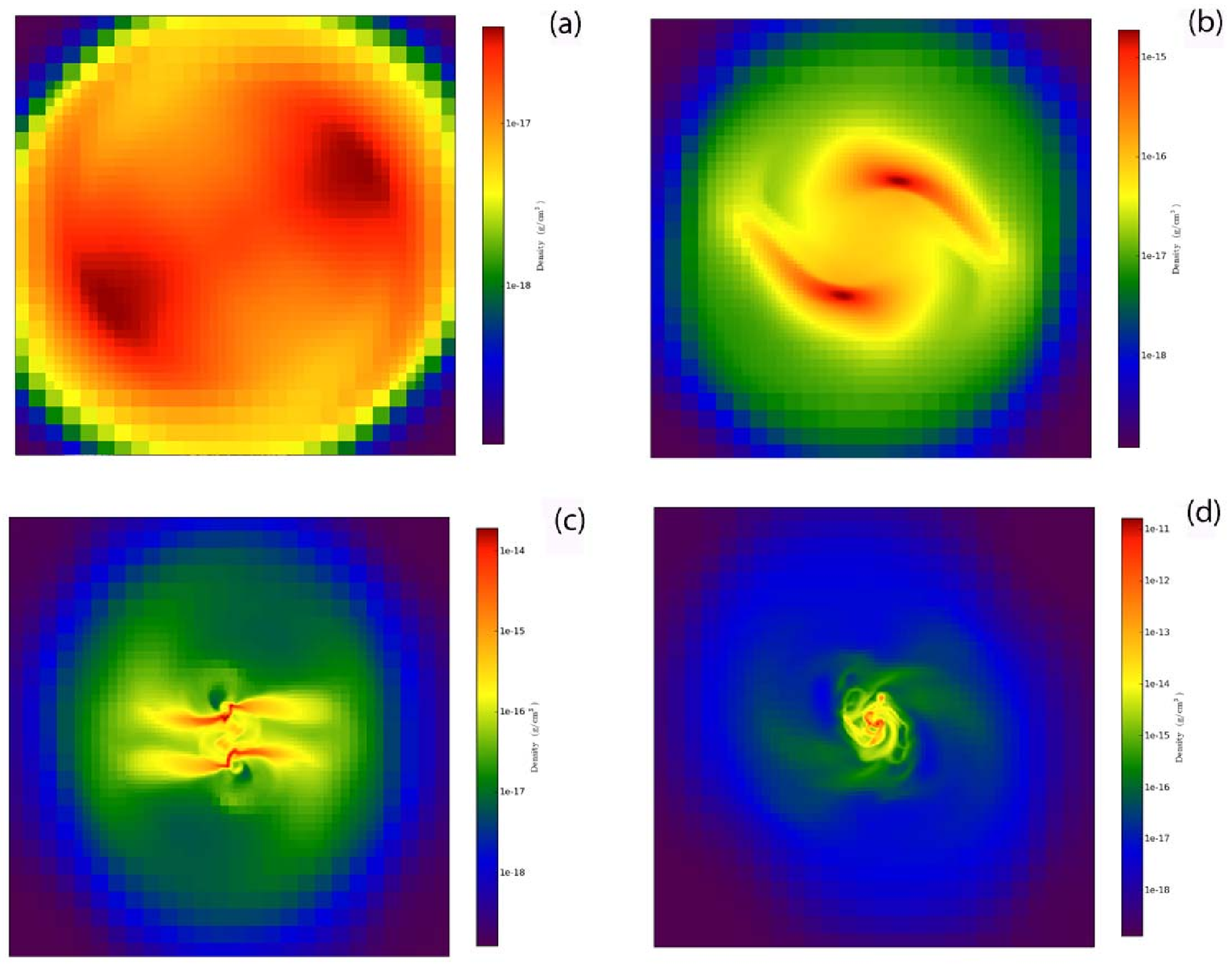}
\caption{Evolution of the midplane density distribution for the isothermal, 
magnetic model mag-x-300 (a magnetic version of model BB79),
shown at times: a - 0.524 $t_{ff}$, b - 1.243 $t_{ff}$, c - 1.540 $t_{ff}$, 
and d - 1.791 $t_{ff}$, plotted as in Figure 1.}
\end{figure}

\clearpage

\begin{figure}
\vspace{-2.0in}
\plotone{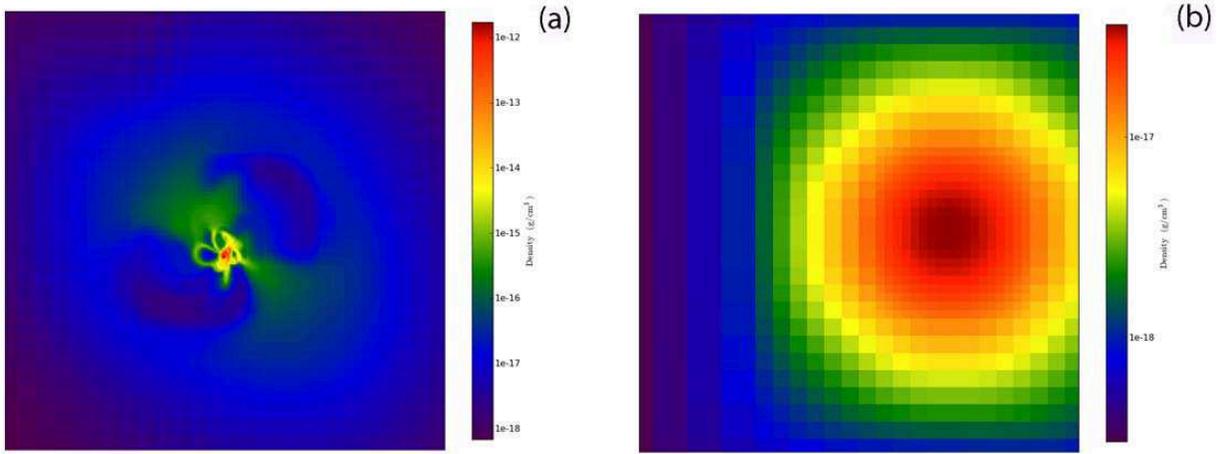}
\caption{Midplane density distributions for models mag-x-800 (a) and 
mag-x-1000 (b), at times of 8.769 and 20.73 $t_{ff}$, respectively, plotted
as in Figure 1.}
\end{figure}

\clearpage

\begin{figure}
\vspace{-2.0in}
\plotone{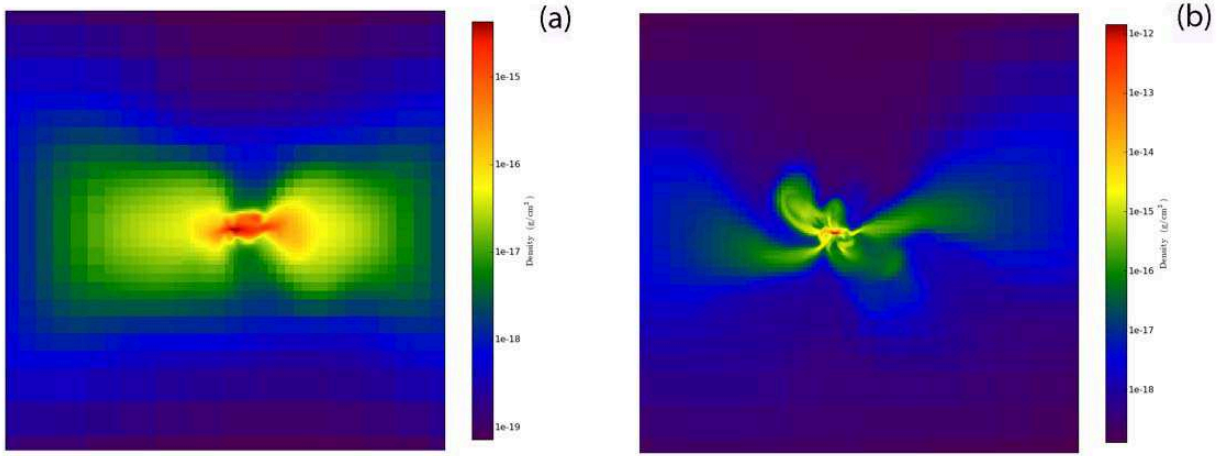}
\caption{Vertical ($x = 0$ plane) density distributions for models 
mag-z-1600 (a) and mag-x-800 (b), at times of 7.742 and 8.769 $t_{ff}$, respectively, plotted as in Figure 1.}
\end{figure}
\clearpage

\begin{figure}
\vspace{-2.0in}
\plotone{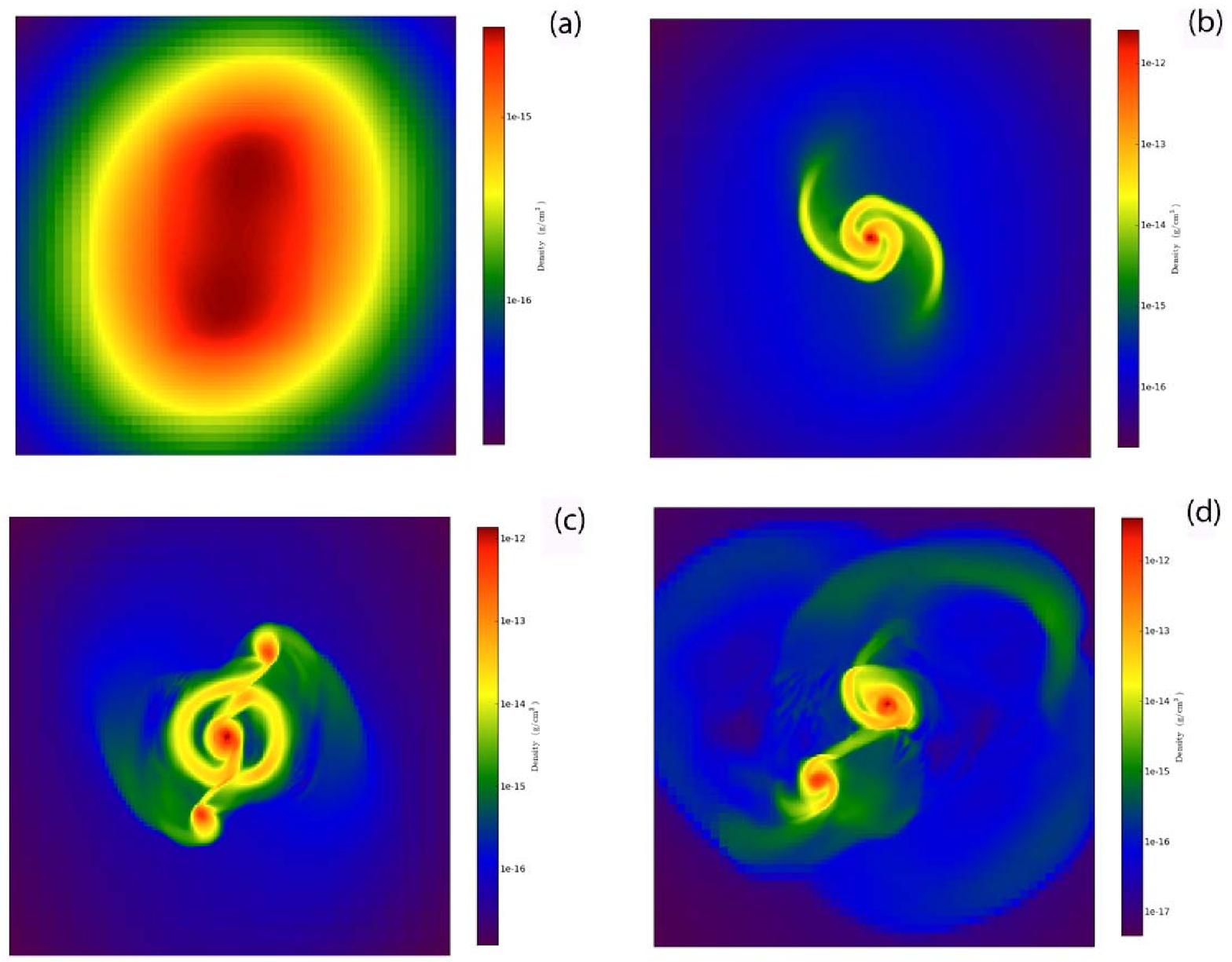}
\caption{Evolution of the midplane density distribution for the barotropic, 
magnetic model poly-z-20 (a magnetic version of model BB79-10),
shown at times: a - 1.179 $t_{ff}$, b - 1.503 $t_{ff}$, c - 1.797 $t_{ff}$, 
and d - 2.200 $t_{ff}$. Region shown is $3.2 \times 10^{16}$ cm across 
at all times.}
\end{figure}

\clearpage

\begin{figure}
\vspace{-2.0in}
\plotone{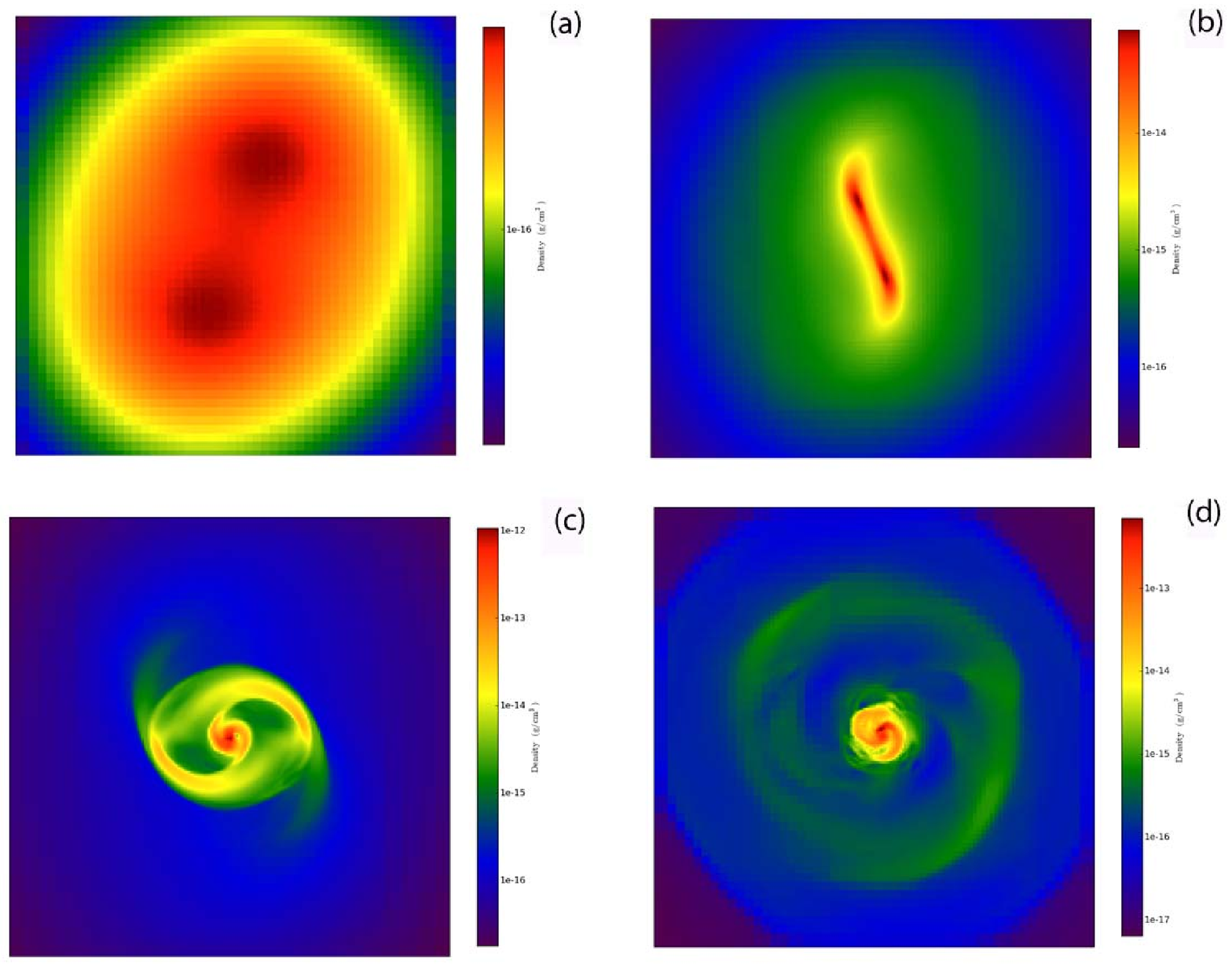}
\caption{Evolution of the midplane density distribution for the barotropic, 
magnetic model poly-z-81 (a magnetic version of model BB79-10),
shown at times: a - 1.089 $t_{ff}$, b - 1.362 $t_{ff}$, c - 1.628 $t_{ff}$, 
and d - 2.200 $t_{ff}$. Region shown is $3.2 \times 10^{16}$ cm across 
at all times.}
\end{figure}

\end{document}